\documentclass[a4, journal]{IEEEtran}
\IEEEoverridecommandlockouts
\usepackage{cite}
\usepackage[hyphens]{url}

\usepackage{amsmath,amssymb,amsfonts}

\newcommand{\mycomment}[1]{}

\usepackage{graphicx}
\usepackage{textcomp}
\usepackage{xcolor}
\usepackage{subfigure}
\usepackage{multirow}
\usepackage{tabularx}
\usepackage{gensymb}
\usepackage{svg}
\usepackage{optidef}
\usepackage[ruled,vlined]{algorithm2e}
\usepackage{hyperref}
\newcommand{\linebreakand}{%
  \end{@IEEEauthorhalign}
  \hfill\mbox{}\par
  \mbox{}\hfill\begin{@IEEEauthorhalign}
}
\usepackage{nomencl}
\usepackage{etoolbox}
\def\BibTeX{{\rm B\kern-.05em{\sc i\kern-.025em b}\kern-.08em
    T\kern-.1667em\lower.7ex\hbox{E}\kern-.125emX}}
    
\renewcommand\nomgroup[1]{%
  \item[\bfseries
  \ifstrequal{#1}{A}{Sets}{%
  \ifstrequal{#1}{B}{Loads and Generation}{%
  \ifstrequal{#1}{C}{Storage}{%
  \ifstrequal{#1}{D}{Medium- and Heavy-Duty (MHD) BEV}{%
  \ifstrequal{#1}{E}{Investment}{}}}}}%
]}

\makenomenclature
\begin{document}

\abovedisplayskip=1pt
\belowdisplayskip=1pt
\abovedisplayshortskip=1pt
\belowdisplayshortskip=1pt

\title{Impact of Flexible and Bidirectional Charging in Medium- and Heavy-Duty Trucks on California's Decarbonization Pathway \\
}

\author{Osten~Anderson,~\IEEEmembership{Student~Member,~IEEE,}
        Wanshi~Hong,~\IEEEmembership{Member,~IEEE}
        Bin~Wang,~\IEEEmembership{Senior~Member,~IEEE}
        and Nanpeng~Yu,~\IEEEmembership{Senior~Member,~IEEE}
}

\maketitle
\IEEEpubidadjcol

\begin{abstract}
California has committed to ambitious decarbonization targets across multiple sectors, including decarbonizing the electrical grid by 2045. In addition, the medium- and heavy-duty truck fleets are expected to see rapid electrification over the next two decades. Considering these two pathways in tandem is critical for ensuring cost optimality and reliable power system operation. In particular, we examine the potential cost savings of electrical generation infrastructure by enabling flexible charging and bidirectional charging for these trucks. We also examine costs adjacent to enabling these services, such as charger upgrades and battery degradation. 
We deploy a large mixed-integer decarbonization planning model to quantify the costs associated with the electric generation decarbonization pathway. Example scenarios governing truck driving and charging behaviors are implemented to reveal the sensitivity of temporal driving patterns. 
Our experiments show that cost savings on the order of multiple billions of dollars are possible by enabling flexible and bidirectional charging in medium- and heavy-duty trucks in California. 
\end{abstract}


\printnomenclature

\section{Introduction}
With the rapid shift towards renewable energy generation as a response to climate change, power system planning has become increasingly important. 
California has set ambitious decarbonization goals across multiple sectors, including transportation with the California Air Resources Board's Advanced Clean Fleet regulation, and electric power generation with Senate Bill 100 and Senate Bill 350. As a result, these sectors are expected to change rapidly over the next two decades. It is crucial that these transitions be planned in tandem to ensure cost-effectiveness and reliable power system operation. 

It is well established that transportation and energy generation are becoming increasingly linked fields as part of the response to climate change.
Transportation electrification is a key component of the energy transition, and vehicle charging load is expected to become a large share of the energy demand as penetration of electric vehicles increases. This transition is occurring together with the shift from carbon-based to renewables-based power generation. 

Adoption of battery electric vehicles (BEV) is expected to both increase electricity demand as well as impact the load patterns. 
A topic of recent interest has been leveraging the charging flexibility of BEVs to reduce the impacts on power grid operation. A major component of this is flexible charging, or V1G, which is the ability to control vehicle charging, typically to shift charging from a peak time to an off-peak time to lower stress on the grid or to adjust the charging power with respect to pricing and demand response signals from the electric utilities. Even further is V2G or bi-directional charging. In this case, vehicles can discharge to the grid, to provide energy shifting or ancillary services. 

It is easier to implement flexible and bidirectional charging in large-scale for medium- and heavy-duty (MHD) BEV than for light duty (LD) BEVs. The number of MHD BEVs is projected to be much fewer; in 2035 the projected LD BEV stock in California is over 15 million, whereas the MHD BEV is approximately 400,000. The smaller number of MHD BEVs and chargers makes it inherently easier to control and coordinate. Simultaneously, MHD BEVs are associated with larger battery capacities than LD BEVs. MHD BEV are also likely to be operated with more sophisticated planning in fleets, and may be less likely to be affected by the randomness of the driving behaviors. MHD BEV adoption may also be shifted towards larger logistics companies with the capital to purchase these vehicles, and operating a large number of vehicles may influence the incentives of enrolling in flexible charging or V2G operations. These unique characteristics make MHD BEVs a more enticing candidate than LD BEVs for pursuing flexible charging and V2G operations.

In this study, the adoption of MHD BEVs is considered exogenous to the decarbonization planning problem. It is assumed that the MHD BEV stock over years aligns with the existing California policy requirements, such as CARB's Advanced Clean Fleet Regulation \cite{CARB}. As such, enabling V1G or V2G services could help avoid installing additional renewable generation or storage capacity with relatively little added cost and difficulty.

%\section{Related Work} \label{sec2}

A great deal of literature has focused on the economic benefit of V1G and V2G. However, the majority of these works have focused on short-term costs and the economic benefit to the BEV owner. 
In \cite{CA_rev}, the potential revenue for BEV owners in California is examined while paying attention to future grid behavior, including wide adoption of BEVs and future grid changes. 
The value of BEVs has been examined for both managing load, including V2G \cite{stochasticV2G} and peak shaving \cite{peak_shaving} and for providing ancillary services, like frequency regulation \cite{EV_freqreg}. The authors in \cite{value_streams} look at a range of potential value streams for V1G and V2G services. 

Somewhat less work has been done to quantify the economic benefit of enabling V1G and V2G services from the perspective of power system planning. These works generally optimize investment planning alongside dispatch and BEV charging scheduling to provide lower infrastructure costs and avoid buildout of generation and energy storage capacity. Ramirez et al \cite{Co-op1} present a co-optimization of power system planning with dispatch of flexible charging with LD BEVs with a UK-based test system. Yao et al \cite{Co-op2}, Suski et al \cite{Co-op3}, Hajebrahimi et al \cite{Co-op4}, and Gunkel et al \cite{Co-op5} present similar co-optimizations with case studies in China, the Maldives, Canada, and the EU, respectively. In \cite{E3V2G}, an analysis of the potential savings of V1G and V2G, including ancillary services, is analyzed for LD BEVs in California. Xu et al \cite{V2GGHG} look at the potential emissions reductions of these services, including life-cycle analysis of BEVs. 

Similarly, most works have focused on LD BEVs, rather than MHD BEV. As discussed above, these groups have rather distinct behavior, which affects both their theoretical value and practical implementability. In this work, we focus on this gap in the literature, and examine the value of V1G and V2G in California's decarbonization pathway, specifically with respect to electrification of medium- and heavy-duty vehicles.
%In general, LDEV spend more time stationary, but are likely to have less defined schedules than fleet-managed MHD BEVs. The population of LDEVs are orders of magnitude greater. Thus, there are more devices to control, with a smaller incremental demand being controlled. This poses additional challenges to the practical implementation of V1G and V2G for LDEVs. 

In this paper, we examine the potential savings and implicit costs of V1G and V2G services through the lens of California's energy transition investment planning. We develop a mixed-integer linear program (MILP) decarbonization planning model incorporating a clustered representation of MHD BEV based on the timing of charging and driving. 
A surrogate Lagrangian relaxation-based technique is implemented to provide computational tractability of the large MILP model.
We analyze the results of the three charging regimes under two MHD BEV driving scenarios, and show a range of potential savings as high as 16 billion dollars. We also examine some of the costs related to charging services to show that the cost savings these services provide are robust. 
The remainder of the paper will be organized as follows. 
%Section \ref{sec2} will review existing works examining the value of V1G and V2G. 
Section \ref{sec3} will formulate the power system planning model. Section \ref{sec4} will describe the methodology used to solve the model. Section \ref{sec5} will discuss results and policy implications. Section \ref{sec6} will present the conclusions. 

\section{Technical Method} \label{sec3}

\nomenclature[A, 01]{$t, T$}{Index, set of hour}
\nomenclature[A, 01]{$w, W$}{Index, set of representative period}
\nomenclature[A, 01]{$y, Y$}{Index, set of year}
\nomenclature[A, 02]{$u, U$}{Index, set of thermal unit}
\nomenclature[A, 03]{$s, S$}{Index, set of storage resource}
\nomenclature[A, 04]{$r, R$}{Index, set of renewable resource}
\nomenclature[A, 05]{$h, H$}{Index, set of large hydro resource}
\nomenclature[A, 06]{$z, Z$}{Index, set of balancing authority zone}
\nomenclature[A, 07]{$l, L$}{Index, set of line}
\nomenclature[A, 07]{$e \in E$}{Index, set of battery electric vehicle (BEV) cluster}

\nomenclature[A, 15]{$U_z$}{Subset of thermal resources in zone $z$}
\nomenclature[A, 16]{$S_z$}{Subset of storage resources in zone $z$}
\nomenclature[A, 17]{$R_z$}{Subset of renewable resources in zone $z$}
\nomenclature[A, 18]{$H_z$}{Subset of large hydro resources in zone $z$}

\nomenclature[B, 01]{$\mathcal{L}_z(t)$}{Load in zone $z$ at time $t$ (MW)}
\nomenclature[B, 02]{$v_{u}(t)$}{On/off status of unit $u$ at time $t$ (1, 0) }
\nomenclature[B, 03]{$p_{u}(t)$}{Power output of unit $u$ at time $t$ (MW)}
\nomenclature[B, 04]{$p_r(t)$}{Power output of renewable resource $r$ at time $t$ (MW)}
\nomenclature[B, 05]{$p_h(t)$}{Power output of large hydro resource $h$ at time $t$ (MW)}
\nomenclature[B, 18]{$f_l(t)$}{Flow on line $l$ at time $t$ (MW)}
\nomenclature[B, 18.5]{$\lambda_{l, z}$}{Incidency of line $l$ on zone $z$}
\nomenclature[B, 21]{$PF_r(t)$}{Production factor of renewable resource $r$ at time $t$}
\nomenclature[B, 22]{$p_r^{curt}$}{Curtailment of renewable resource $r$ (MW)}
\nomenclature[B, 23]{${c^{curt}_r}$}{Cost of curtailment of resource $r$ (\$/MWh)}

\nomenclature[B, 24]{$SUC_u(t)$}{Startup cost of unit $u$ at time $t$ (\$)}
\nomenclature[B, 25]{$SDC_u(t)$}{Shutdown cost of unit $u$ at time $t$ (\$)}
\nomenclature[B, 26]{$GCS_u$}{Generation cost slope of unit $u$ (\$/MWh)}
\nomenclature[B, 27]{$GCI_u$}{Generation cost intercept of unit $u$ (\$/hour)}

%storage params
\nomenclature[C, 01]{$v_s(t)$}{Storage charge (0)/discharge (1) status at time $t$}
\nomenclature[C, 02]{$p_{s}^c(t)$}{Storage rate of charge at time $t$ (MW)}
\nomenclature[C, 03]{$p_s^d(t)$}{Storage rate of discharge at time $t$ (MW)}
%\nomenclature[C, 04]{$\overline{p}_s^{c}$}{Storage max rate of charge (MW)}
%\nomenclature[C, 05]{$\overline{p}_s^{d}$}{Storage max rate of discharge (MW)}
%\nomenclature[C, 06]{$\overline{C}_s$}{Storage max state of charge (MWh)}
%\nomenclature[C, 07]{$\underline{C}_s^$}{Storage min state of charge (MWh)}
\nomenclature[C, 08]{$C_s(t)$}{Storage state of charge at time $t$ (MWh)}
\nomenclature[C, 09]{$\eta_s^c$}{Storage charge efficiency}
\nomenclature[C, 10]{$\eta_s^d$}{Storage discharge efficiency}
\nomenclature[C, 11]{$\delta_s$}{Storage self discharge}

%EV params
\nomenclature[D, 0]{$v_e(t)$}{MHD BEV charge (0)/discharge (1) status at time $t$}
\nomenclature[D, 01]{$p_{e}^c(t)$}{MHD BEV charge at time $t$ (MW)}
\nomenclature[D, 02]{$p_{e}^d(t)$}{MHD BEV discharge at time $t$ (MW)}
\nomenclature[D, 03]{$\overline{P}_{e}$}{MHD BEV charger power rating (MW)}
\nomenclature[D, 04]{$C_{e}(t)$}{MHD BEV state of charge at time $t$ (MWh)}
\nomenclature[D, 05]{$\overline{C}_{e}$}{MHD BEV maximum state of charge (MWh)}
\nomenclature[D, 06]{$\underline{C}_{e}$}{MHD BEV minimum state of charge (MWh)}
\nomenclature[D, 08]{$t^{depot}_e$}{Hour of depot arrival}
\nomenclature[D, 09]{$t^{drive}_e$}{Hour of drive start}
\nomenclature[D, 10]{$C^{depot}_e$}{State of charge at depot arrival (MWh)}
\nomenclature[D, 11]{$C^{drive}_e$}{State of charge at drive start (MWh)}

%investment params
\nomenclature[E, 01]{$IU_u(y)$}{Install status of unit $u$ in year $y$}
\nomenclature[E, 02]{$IU^p_u(y)$}{Planned install status of unit $u$ in year $y$}
\nomenclature[E, 03]{$IU^b_u(y)$}{Build flag for unit $u$ in year $y$}
\nomenclature[E, 04]{$IU^r_u(y)$}{Retirement flag for unit $u$ in year $y$}
\nomenclature[E, 05]{$IC_s(y)$}{Installed capacity of storage $s$ in year $y$ (MW)}
\nomenclature[E, 06]{$IC^p_s(y)$}{Planned capacity of storage $s$ in year $y$ (MW)}
\nomenclature[E, 07]{$IC^b_s(y)$}{Built capacity of storage $s$ in year $y$ (MW)}
\nomenclature[E, 08]{$ICE_s(y)$}{Installed energy capacity of storage $s$ in year $y$ (MWh)}
\nomenclature[E, 09]{$ICE^p_s(y)$}{Planned energy capacity of storage $s$ in year $y$ (MWh)}
\nomenclature[E, 10]{$ICE^b_s(y)$}{Built energy capacity of storage $s$ in year $y$ (MWh)}

\nomenclature[E, 11]{$IC_r(y)$}{Installed capacity of renewable $r$ in year $y$ (MW)}
\nomenclature[E, 12]{$IC^p_r(y)$}{Planned capacity of renewable $r$ in year $y$ (MW)}
\nomenclature[E, 13]{$IC^b_r(y)$}{Built capacity of renewable $r$ in year $y$ (MW)}

\nomenclature[E, 14]{$C_y^{gen}$}{Generation costs in year  $y$ (\$)}
\nomenclature[E, 15]{$C_y^m$}{Maintenance costs in year  $y$ (\$)}
\nomenclature[E, 16]{$C_y^{inv}$}{Investment costs in year  $y$ (\$)}

In this section, we formulate the planning problem as an integer linear program optimization over two timescales. Unit commitment and economic dispatch is modeled hourly by scheduling generation to satisfy load and ancillary service requirements. Investment is modeled at the yearly timescale, and governs the construction and retirement of energy resources. The two timescales are linked by several constraints, including policy constraints like emissions limits, and constraints governing the operation of resources based on their investment status. Subsection \ref{sec3A} formulates the hourly unit commitment model. Subsection \ref{sec3B} then integrates the unit commitment and dispatch model into the planning model.

\subsection{Unit Commitment} \label{sec3A}

Unit commitment (UC) is modeled at hourly frequency over a set of hours $T$. 
In general, unit commitment variables are indexed temporally by a tuple $(y,w,t)$ of year, week, and time. However, when describing unit commitment alone, only the last index is relevant, because we do not yet consider investment and policy constraints which link weeks and years. For notational brevity, we will hide the axis of year and week. That is, for this subsection, $p_u(y,w,t) \rightarrow p_u(t)$ for an arbitrary $y,w$. Finally, we will define the set of all generation and transmission constraints discussed in this section as $\Omega$.

Representative periods $w$ are treated as circular or cyclical. In essence, the last hour in $T$ links back to the first hour, and all constraints linking hours are enforced accordingly. This cyclical representation is modeled via the modulo operator $\tau(t) = \mod(t-1+T, T)$. For constraints not linking hours, only the regular period is enforced and $\tau(t)=t$.
%As in the RESOLVE package \cite{RESOLVEcode}, a widely used electricity resource planning model, time periods will be considered circular or cyclical. In essence, the last hour in $T$ links back to the first hour. This avoids the complications and computational complexity around defining or solving the initial status of units. All constraints which link hours together are enforced accordingly. This cyclical representation is modeled via the modulo operator. 
%\begin{equation}
%    \tau(t) = \mod(t-1+T, T).
%\end{equation}
%Constraints which extend past $T$ map back to the beginning of the period. Similarly, a negative argument $t$ in $\tau(t)$ maps backwards from the end of the period. In the regular period, $t \in [1, T]$, $\tau$ can be omitted as $t=\tau(t)$.

\subsubsection{Generation Resources}
The generation fleet consists of five classes of generation resources, each with distinct operational characteristics: thermal units, renewable resources, firm resources, storage resources, and large hydro resources. These classes and relevant constraints will be briefly discussed in this section. The full formulation of these generation constraints can be found in \cite{anderson}. Flexible MHD BEV charging is modeled as a demand-side resource, and will be discussed in detail. 

\noindent \textbf{Thermal Units.} 
Thermal units include a variety of combustion-based power plants, such as coal-fired power plants, combined-cycle gas turbines, peakers, steam turbines, and aeroderivative combustion turbines, each with unique technical operating characteristics.
These resources are dispatchable, and the commitment of thermal units is modeled as binary. The output $p_u(t)$ of resource $u$ is thus constrained by minimum and maximum output based on commitment $v_u(t)$:
\begin{equation} \label{gen1}
    \underline{P}_uv_u(t) \leq p_u(t) \leq \overline{P}_uv_u(t), \; \forall u \in U, t \in T,
\end{equation}
where $\underline{P}_u$ and $\overline{P}_u$ are the minimum and maximum power levels for unit $u$. Thermal units are further constrained by minimum uptime and downtime, startup and shutdown limits, and ramp limits. 

\noindent \textbf{Renewable and Firm Resources.} 
Renewable resources consist of solar and wind farms, as well as aggregated behind-the-meter solar photovoltaic systems. Firm resources include nuclear, small hydro, biofuel, geothermal, and combined heat and power. Firm resources are lumped with renewables as they have generally similar properties. Each resource generates according to the product of the installed capacity $IC_r(y)$, at an arbitrary year, and an hourly factor $PF_r(t)$ accounting for meteorological conditions associated with renewable generation, less any curtailment. Thus, the generation of these resources is given by: 
\begin{equation} \label{renpower}
    p_r(t) = IC_r \cdot PF_r(t) - p_r^{curt}(t), \; \forall r \in R, t \in T.
\end{equation}
The power output of firm resources is not subject to hourly fluctuations. However, it can experience seasonal variations, such as maintenance-related changes for nuclear power or changes in stream flow for small hydroelectric systems. Still, they can be associated with an hourly capacity factor. Firm resources are not curtailable, so $p_r^{curt}(t) = 0$ for those resources. 

\noindent \textbf{Large Hydro Units.} Unlike small hydro, large hydro units are dispatchable hydropower resources. The output of large hydro units $p_h(t)$ is subject to ramp limits, minimum and maximum output constraints, and an energy budget constraint. 

\noindent \textbf{Storage Resources.} 
Storage resources include both pumped hydro storage and stationary battery storage. These resources are modeled using a binary indicator of charge or discharge status in order to enforce minimum charge/discharge duration constraints for the case of pumped hydro. This binary also prevents simultaneous discharge and charge. These resources are characterized by their power capacity (MW) and energy capacity (MWh). These resources are accordingly subject to charge and discharge limits and state of charge (SoC) limits. It is necessary to track the state of charge for these units, including losses due to efficiency:
\begin{multline} \label{SoC}
C_{s}(t) = [(1-v_{s}) p_{s}^c(t) \eta_{s}^c - v_{s} p_{s}^d(t) \frac{1}{\eta_{s}^d}] \times 1 \: hour + \\(1-\delta_{s}) C_{s}(\tau(t-1)),  \forall s \in S, t \in T
\end{multline}
\noindent \textbf{Flexible MHD BEV Charging.}
MHD BEV with flexible or bidirectional charging capability, is modeled similarly to storage resources, with the major exception that a large amount of discharge happens exogenously through driving, during which these resources are not connected to the power grid. 
To integrate MHD BEV into the planning framework, these resources are modeled as dispatchable by a central system operator, rather than a virtual power plant controlled by price signals. 
Each vehicle is associated with a charge start time, charge end time, starting state-of-charge, and ending state-of-charge. It is assumed that the vehicle is plugged in and available for charging for the entire duration that it is at the depot. These values essentially determine the vehicles charging needs, as well as potential V2G provisions. Modeling vehicles individually would make the problem computationally intractable; thus, vehicles are grouped by their start and end hour to form virtual power plants. The power and energy capacity parameters of the clusters are obtained as the summation of the individual parameters of the MHD BEVs in the cluster. MHD BEVs are modeled as a demand-side resource. 

The control of MHD BEV clusters within optimization is operationalized by three variables: state of charge $C_e(t)$, charge power $p_e^c(t)$, and discharge power $p_e^d(t)$. These three variables are subject to limits based on the capacity of the cluster, as well as the timing at which the cluster is connected to the grid at the depot for charging vs disconnected from the grid for driving.
Discharge through driving is exogenous, and $p_e(t)=0$ when the vehicle is not at the depot. If V2G is not considered, discharge is not allowed and $p_e^d(t) = 0, \; \forall t \in T$.

The definition of a period $T$ allows for multiple days to be modeled consecutively, and the same charge events occur each day. To account for this, we define the set of days in the period $D$, where $|D|=|T|/24$ denotes the number of days in the period. We also define a time wrap $t_e^\Delta$, to account for charging which occurs overnight. For each day, the variable state of charge at the time of depot arrival and departure is set equal to the input state of charge at the start \eqref{eq:ev_start_ch} and end of charging \eqref{eq:ev_end_ch}. 
\begin{equation} \label{eq:ev_start_ch}
    C_{e}(t^{depot}_e + d\cdot24) = C^{depot}_e, \forall d \in D, e \in E
\end{equation}
\begin{equation} \label{eq:ev_end_ch}
    C_{e}(t^{drive}_e + d\cdot24) = C^{drive}_e, \forall d \in D, e \in E
\end{equation}
\begin{equation} \label{eq:timewrap}
    t^\Delta_e = 24\text{ if }t^{depot}_e > t^{drive}_e\text{ else }0
\end{equation}
While the vehicle is at the depot, bounds of charge \eqref{eq:evchargepower} and discharge rate \eqref{eq:evdischargepower}, and bounds on state of charge are enforced \eqref{eq:evsoc}. State of charge is also tracked with provisions for charger efficiency \eqref{eq:evsoctrack}. 
%\begin{multline} \label{eq:evpower}
%    \underline{P}_{e} <= p_e(\tau(t+d\cdot24)) <= \overline{P}_{e}, \\ \forall e \in E, d \in D, t \in [t^{depot}_e, t^{drive}_e+t^\Delta_e]
%\end{multline}
\begin{multline} \label{eq:evchargepower}
    0 <= p_e^c(\tau(t+d\cdot24)) <= (1-v_e(t))\overline{P}_{e}, \\ \forall e \in E, d \in D, t \in [t^{depot}_e, t^{drive}_e+t^\Delta_e]
\end{multline}
\begin{multline} \label{eq:evdischargepower}
    0 <= p_e^d(\tau(t+d\cdot24)) <= v_e(t)\overline{P}_{e}, \\ \forall e \in E, d \in D, t \in [t^{depot}_e, t^{drive}_e+t^\Delta_e]
\end{multline}
\begin{multline} \label{eq:evsoc}
    \underline{C}_e <= C_{e}(\tau(t)) <= \overline{C}_e, \\ \forall e \in E, d \in D, t \in [t^{depot}_e, t^{drive}_e+t^\Delta_e]
\end{multline}
\begin{multline} \label{eq:evsoctrack}
    C_e(\tau(t+1) = C_e(\tau(t)) + p_e^c(\tau(t) \eta_e^c - p_e^d(\tau(t)) \eta_e^d, \\ \forall e \in E, d \in D, t \in [t^{depot}_e, t^{drive}_e+t^\Delta_e-1]
\end{multline}
\subsubsection{Zones and Lines}
A zonal unit commitment model is used to represent the Western Interconnection. The model is directed towards California's decarbonizaton goals, and this is reflected in the zonal modeling. The formulation presented here is easily adaptable to other zones.
As the main balancing authority in California, the model focuses on CAISO, and additionally includes smaller balancing authorities in California (LADWP, IID, BANC). Balancing authorities in the Northwest and Southwest are represented by two aggregations. 
Transmission corridors between zones are represented using a transport model, in which the line flows $f_l(t)$ is a decision variable. This approach greatly simplifies the computational complexity associated with an optimal power flow formulation, while still effectively representing the system interconnections. Line flow can be positive or negative, where negative flow means flow opposite of the line reference direction. Each line is associated with two zones and a reference direction, and this is encoded in $\lambda_{l,z}$. If line $l$ is not incident on node $z$ then $\lambda_{l,z}$ is 0, and $\lambda_{l,z}$ is $+1$ or $-1$ if $l$ goes to or from $z$, respectively. Line flows are additionally subject to transmission line limits.

\subsubsection{Load and Reserve Requirements} 
Zonal power balance constraints ensure that the generation and net line flows meet the load. Each zone must satisfy these constraints as:  
\begin{flalign} \label{Nodalpowerbalance}
& \sum_{u \in U_z} p_i(t) + \sum_{s \in S_z}[p_{s}^d(t) - p_{s}^c(t) ] + \sum_{r \in R_z } p_r(t) + \sum_{h \in H_z} p_h(t) \nonumber\\ &  + \sum_{l \in L} \lambda_{l,z} f_l(t) = \mathcal{L}_z(t) + \sum_{e \in E_z}[p_e^c(t) - p_e^d(t)], \;% t \in T, z \in Z.
\end{flalign}
CAISO must additionally satisfy the requirement for ancillary services. These products ensure reliable grid operation, and include frequency response, spinning reserve, regulation up and down, and load following up and down. 

\subsubsection{Unit Commitment Objective}
The unit commitment objective function \eqref{ucobj} is to minimize the total cost of fuel, startup and shutdown, power transmission, and renewable curtailment:
\begin{equation} \label{ucobj}
    \min \mathcal{C}^{gen}
\end{equation}
\begin{flalign} \label{cgen(yw)}
\mathcal{C}^{gen} & = \sum_{t \in T} \sum_{u \in U}
\Big\{SUC_u(t)+SDC_u(t) \nonumber \\ & + (GCI_u \cdot v_u(t) + GCS_u \cdot p_u(t)) \times 1 \: hour \Big\} \\  +  [\sum_{t\in T} &\sum_{l \in L} f_l(t) \cdot \emph{c}^{tx}_l + \sum_{t\in T} \sum_{r \in R} \emph{c}^{curt}_r \cdot p_r^{curt}(t)] \times {1 \:  hour}. \nonumber
\end{flalign}
%\begin{equation} \label{cgen(yw)}
%\begin{split}
%where \;  \mathcal{C}^{gen} & = \sum_{t \in T} \sum_{u \in U}
%\Big\{SUC_u(t)+SDC_u(t) \\ & + (GCI_u \cdot v_u(t) + GCS_u \cdot p_u(t)) \times 1 \: hour \Big\} \\  +  [\sum_{t\in T} \sum_{l \in L}  & f_l(t) \cdot \emph{c}^{tx}_l + \sum_{t\in T} \sum_{r \in R} \emph{c}^{curt}_r \cdot p_r^{curt}(t)] \times {1 \:  hour}.
%\end{split} 
%\end{equation}
\subsection{Decarbonization Planning} \label{sec3B}
Decarbonization planning folds the unit commitment formulation into a multi-year model allowing for the build and retirement of resources while enforcing policy constraints related to emissions and renewable generation. The goal is to identify the investment strategy which, while meeting all constraints, minimizes the cost of energy generation, fleet maintenance, and capital costs of constructing new capacity.

The present study focuses on California's decarbonization goals, so development of new resources is restricted to CAISO. Addition of capacity to match with load growth in other zones is exogenous. However, the formulation described here is applicable to multi-zone investment. Candidate resources include various wind, solar, energy storage, geothermal, and biomass projects, as well as new lower emission power plants. Economic retirement is also available for existing thermal power plants. 

First, we formulate the investment variables of each resource class, and demonstrate how investment interfaces with dispatch constraints. For thermal units, $IU_u(y)$ is a binary indicator of a unit being operational (1) or not (0). New construction and retirement are modeled separately, with $IU_u^b(y)=1$ if the unit is built in year $y$ and $IU_u^r(y)=1$ if the unit is retired. Unit commitment interfaces with investment by \eqref{OUV}, which only requires units be operational to be committed. 
\begin{equation} \label{IC}
\begin{split}
    IU_u(y) = IU_u^p(y) + \sum_{\mathcal{Y}=1}^{y} ( IU_u^b(\mathcal{Y}) - IU_u^r(\mathcal{Y}))%, \forall y \in Y
\end{split}
\end{equation}
\begin{equation} \label{OUV}
\begin{split}
    IU_u(y) \geq v_{u}(y,w,t),\; \forall u \in U,\: w \in W,\: t \in T
\end{split}
\end{equation}
Additional capacity of renewables and storage can be installed as continuous variables. For storage, the capacity of storage energy and storage power are modeled separately, as they constitute different pieces of hardware (inverters and battery cells). 
\begin{flalign} \label{ICE}
    IC_s(y) = IC_s^p(y) + \sum_{\mathcal{Y}=1}^{y} ( IC_s^b(\mathcal{Y}) - IC_s^r(\mathcal{Y}))\\
    ICE_s(y) = ICE_s^p(y) + \sum_{\mathcal{Y}=1}^{y} ( ICE_s^b(\mathcal{Y}) - ICE_s^r(\mathcal{Y}))
\end{flalign}
Storage investment interfaces with dispatch through the bounds on SoC, charge, and discharge. The maximum rate of charge and discharge is equal to the installed capacity: $\overline{p}_s^{c}(y) = \overline{p}_s^d(y) = IC_s(y)$. Typically for stationary battery energy storage, the full energy capacity is not utilized and some headroom/footroom is reserved to lower degradation; for pumped storage, this is not a concern. Thus, the maximum capacity is represented as $\overline{C}_s(y)=ICE_s(y) \cdot \epsilon^{max}_s$, with the minimum represented in a similar way.

Installed capacity of renewable resources is defined in the same way. Investment interfaces via $IC_r(y)$ in \eqref{renpower}.
\begin{equation} \label{ICR}
\begin{split}
    IC_r(y) = IC_r^p(y) + \sum_{\mathcal{Y}=1}^{y} ( IC_r^b(\mathcal{Y}) - IC_r^r(\mathcal{Y})).
\end{split}
\end{equation}
The decarbonization aspect of the present planning problem is operationalized by yearly constraints on emissions and constraints on the percentage of energy served by renewable resources. As the focus of this study is California, policy constraints are only enforced within CAISO. Thus we specify $z=0$ corresponding to CAISO. The emissions of all CAISO thermal units as well as emissions associated with imported energy are subject to the emissions limit. The emissions of thermal unit operation are accounted for in a similar way to fuel cost, via emission slope and intercept terms $e_u^s$ and $e_u^i$. Import emissions are assigned a constant ton/MWh rate $e_l$. Only imports are considered for the emissions constraints, and net exports do not serve to lower the total emissions. 
\begin{multline} \label{emissions}
    E_y \geq \sum_{w \in W} \omega_w \cdot \sum_{t \in T} \biggl(\sum_{u \in U_{z}} e_u^s \cdot p_u(y,w,t) + e_u^i \cdot v_u(y,w,t)\\ + \sum_{l \in L} e_l \cdot \max (0, \lambda_{l,z}f_l(y, w, t))  \biggr)%, y \in Y, z=0.
\end{multline}
The renewable portfolio standard (RPS) constraint requires that a certain fraction $RPS_y$ of the total load each year come from renewable sources. Curtailed renewable energy does not count. The eligibility of each resource in $R$ is given by binary $RPS^{eligible}_r$, as resources like nuclear are grouped with renewables $R$ but do not count towards this constraint.
\begin{multline} \label{RPS}
   RPS_y \cdot \sum_{w \in W} \sum_{t \in T} \mathcal{L}_z(y, w, t) \leq \\ 
   \sum_{w \in W} \omega_w \cdot \sum_{r \in R} \sum_{t \in T} p_r(y,w,t) \cdot RPS^{eligible}_r
\end{multline}
The planning reserve margin (PRM) is a policy constraint directed towards reliability rather than decarbonization. The PRM ensures enough total capacity is held to meet the forecasted peak load with some additional headroom. Each resource class contributes towards the PRM by a fraction of its capacity. Net qualifying capacity is used for thermal resources, a fraction typically close to 1, as well as large hydro. Intermittent energy resources like wind and solar have a more complex relationship. The cumulative contribution is given by the effective load-carrying capacity $ELCC_y$, a 3-dimensional piecewise linear surface in which the contribution declines with increasing penetration of this variable resources. Similarly, $ELCC_{y,s}$ is used for the contribution of storage resources. Specifics of the calculation of ELCC can be found in \cite{anderson}.
\begin{multline} \label{PRM}
 PRM_y \leq \sum^{u \in U_{z}} IU_{y, u} \overline{P}_u NQC_{u} +  ELCC_{y,s} \\ + ELCC_{y} + \sum^{h \in H_{z}} IC_{y, h} NQC_{h}
\end{multline}

As previously mentioned, there are three components of cost that are optimized over: investment, maintenance, and generation. Each modeled year, otherwise referred to as investment interval, is associated with a yearly cost for each component, and these yearly costs are weighted by a yearly weight $\omega_y$.
This yearly weight accounts for the number of years that each investment interval represents and an adjustment for the time value of money from the first year.

The cost of energy generation in year $y$ is $\mathcal{C}_y^{gen}$. Each $y \in Y, w \in W$ is associated with a cost of generation $\mathcal{C}^{gen}(y,w)$ according to \eqref{cgen(yw)}. Due to the intractability in modeling all 8760 hours per year, the year is represented by a subspace consisting of the set of representative periods $W$. Each $w \in W$ is associated with a weight $\omega_w$ encoding the fraction of the year that it represents. These weights are chosen such that $\sum_{w \in W} \omega_w \times ||T|| = 8760$. Consequently, the annual cost of generation is given as: 
\begin{equation} \label{yearlygen}
\begin{split}
    %\mathcal{C}_y^{gen} = \sum^{w \in W} \sum^{t\in T} \sum^{i \in I} \sum^{u \in U} SUC_{i, u}(y, w, t) + SDC_{i, u}(y, w, t) \\ +  GCI_i \cdot v_{i, u}(y, w, t) + GCS_i \cdot p_{i, u}(y, w, t)
    \mathcal{C}_y^{gen} = \omega_y\sum_{w \in W}\omega_w{C}_{y, w}^{gen}.
\end{split}
\end{equation}
Maintenance costs are assessed yearly according to the total operational capacity of each resource. Storage resources have separate cost components for the MWh energy rating $c^m_{s,E}$ and the MW power rating $c^m_{s,P}$. Thermal units are assessed maintenance costs per  unit, and renewable resources have a cost based on the installed MW capacity. Economic retirement allows for maintenance costs to be avoided when capacity is no longer required. 
Then, the cost of maintenance for the year $y$ is: 
\begin{multline}
 \mathcal{C}_y^{m}= \omega_y \biggl(  \sum_{u \in U} IU_{u, y} \cdot c^{m}_i + \sum_{s \in S} ICE_{s, y} \cdot c^{m, E}_{s} + \\ \sum_{s \in S} IC_{s, y} \cdot c^{m, P}_{s} + \sum_{k \in K} IC_{k, y} c^{m}_k + \sum_{h \in H} IC_{h, y} \cdot c^{m}_h \biggr)
\end{multline}
Investment costs are the costs of constructing new resources. The cost for each resource is annualized, and assessed for the duration of the optimization horizon by weighting the annualized cost with the sum of the subsequent $\omega_y$. As before, thermal units have cost per unit $c_{y,u}^{cap}$, renewable resources have cost per MW $c_{y,s}^{cap, P}$, and storage resources have cost per MW $c_{y,r}^{cap}$ and MWh $c_{y,s}^{cap, E}$ separately.
\begin{multline} \label{invcosts}
    \mathcal{C}_y^{inv} =  \biggl(\sum_{u \in U} (IU_u^b(y)) \cdot c_{y,s}^{cap} + \sum_{s \in S} (IC_s^b(y)) \cdot c^{cap, P}_s + \\  + \sum_{s \in S} (ICE_s^b(y)) \cdot c^{cap, E}_s  + \sum_{r \in R} (IC_r^b(y)) \cdot c^{cap}_r \biggr) \cdot \sum_{\gamma=y}^{|Y|} \omega_\gamma
\end{multline}
Finally, the objective for decarbonization planning is to minimize the sum of these yearly costs:
\begin{equation}
    \mathbb{O} = \sum_{y \in Y} \big\{\mathcal{C}_y^{gen} + \mathcal{C}_y^{m} + \mathcal{C}_y^{inv}\big\}.
\end{equation}

\section{Solution Methodology} \label{sec4}
As formulated in the previous section, decarbonization is a mixed-integer linear program. 
\begin{flalign} \label{O}
&\min \;  \{\mathbb{O}\}  \nonumber \\ 
& s.t.\; \Omega\; , \eqref{eq:ev_start_ch}-\eqref{eq:evsoctrack},\; \eqref{Nodalpowerbalance}\; \forall y \in Y, w \in W,\\ & \eqref{IC} - \eqref{ICR}, \eqref{emissions} - \eqref{PRM} \; \forall y \in Y \nonumber 
\end{flalign}
Although commercial MILP solvers improve year over year, they still suffer from the issue of combinatorial complexity. As the number of binary variables increases, the solution time increases superlinearly. Modeling hundreds of thermal units over multiple years each with hundreds of hours quickly becomes impossible to solve in reasonable CPU time. 

To achieve computational tractability, a common approach in the power system planning is to relax the binary variables, often in tandem with clustering thermal units together and approximating their dispatch. This greatly improves the computation time of planning problems, albeit at the loss of model rigor. Instead, we solve the model using surrogate Lagrangian relaxation.

The key to this approach is to relax a difficult constraint, in our case the zonal power balance, and to add the violation of that constraint into the objective function alongside Lagrangian multipliers $\Lambda$. 
The constraint violations of the zonal power balance are given by $r_z(y,w,t) = \big\{\sum_{u \in U_z} p_u(y, w, t) + \sum_{s \in S_z}[ p_{s}^d(y, w, t) - p_{s}^c(y, w, t) ] + \sum_{r \in R_z } p_r(y, w, t) + \sum_{h \in H_z} p_h(y, w, t) + \sum_{l \in L} \lambda_{l,z}f_l(y, w, t) - \mathcal{L}_z(y, w, t)\big\}$.  $\mathbf{R}$ is a vector of constraint violations where $\mathbf{R} = [r_z(y,w,t), \forall z \in Z, y \in Y, w \in W, t \in T]$. 
The resulting optimization is referred to as the dual problem. 
\begin{flalign} \label{lr}
&\min \;  \{\mathbb{O} + \Lambda \cdot R\}  \nonumber \\ 
& s.t. \; \Omega, \eqref{eq:ev_start_ch}-\eqref{eq:evsoctrack}\; \forall y \in Y, w \in W,\\ & \eqref{IC} - \eqref{ICR}, \eqref{emissions} - \eqref{PRM} \; \forall y \in Y \nonumber 
\end{flalign}
The optimization \eqref{lr} is repeatedly solved while updating the multipliers $\Lambda$. 
The dual problem \eqref{lr} is already easier to solve than the one in \eqref{O}. In addition, it becomes possible to solve only a portion of all variables in each iteration while fixing the other variables to their most recent solution, and update the multipliers along the subgradient. This both improves iteration time and improves convergence of $\Lambda$, a known drawback to the traditional Lagrangian relaxation technique. Further details on the implementation of surrogate Lagrangian relaxation, including updating of multipliers $\Lambda$, can be found in \cite{anderson}. 

It is unlikely, and unnecessary, that the multipliers converge until constraint violations are identically zero. Instead, when the multipliers have converged such that the violations are sufficiently low, return to the primal problem \eqref{O} and solve while fixing the majority of the binary variables to the final value in the dual problem. This provides a near-optimal solution to the primal problem in orders-of-magnitude less time than solving the primal problem directly. 

\section{Numerical Study} \label{sec5}

This section will quantify the impact of V1G and V2G on Decarbonization planning.
Subsection \ref{mhdevdata} will introduce two MHD BEV driving and charging datasets and processing them into planning model inputs. Then, we will present the results of the study, both in terms of cost savings and the overall impact on power system investment. 
Finally, we will examine some of the relevant costs, namely battery degradation and charging infrastructure, associated with V1G and V2G to draw conclusions about the value of adopting these services.

%\subsection{Setup for the Planning Model}
The decarbonization model is a zonal representation of the Western Interconnection. The model focuses on CAISO, but also represents 3 small balancing authorities in California (LADWP, BANC, IID) and 2 aggregations of balancing authorities outside California in the Northwest and Southwest. Data is primarily taken from the RESOLVE implementation published by the California Public Utilities Commission \cite{RESOLVEcode}.
Representative periods are selected using the sampling method in \cite{anderson2024selection}. We use 10 representative periods of 3-day length. Investment is modeled in 5-year frequency from 2025 through 2045. Financing is considered through 2065. 

\subsection{Specifications for MHD BEV} \label{mhdevdata}
Accurate modeling of V1G and V2G services requires projections of both the number of MHD BEVs and the operating characteristics of each vehicle, such as drive duration and miles traveled.
In general, there is a great deal of uncertainty associated with long term planning models, due to the reliance on projections of future load, technology costs, and so on. This is compounded by the fact that this planning model is reliant both on the adoption of MHD BEV as well as the usage characteristics. While datasets exist on the driving and parking characteristics of gas and diesel trucks, it is not known if the use cases of MHD BEV will be the same. 

To address this, we examine the impact of V1G and V2G MHD BEVs utilizing the simulated trip patterns in the HEVI-LOAD tool and we build an additional scenario informed by the temporal patterns extracted from a historical truck driving dataset, FleetDNA \cite{fleetdna}. 
%The second scenario is a best-case scenario for the enrollment of vehicles in grid-controlled charging services.
%It is related to the approach presented in \cite{NREL_evpro}. 

\begin{figure}[htb]
\centering
\includegraphics[width=0.9\linewidth]{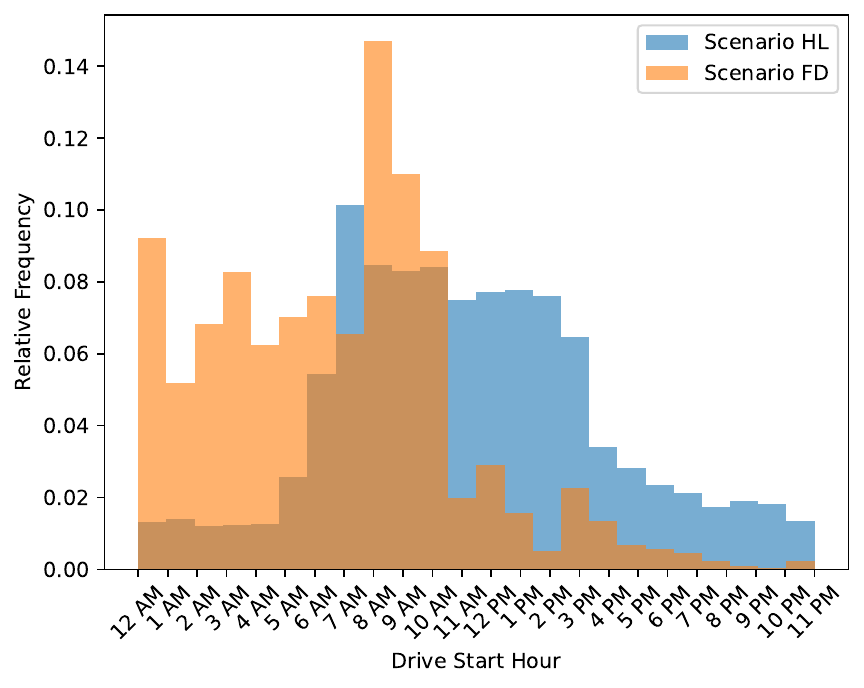}
\caption{Comparison of drive start times between two scenarios.}
\label{fig:scenarios}
\end{figure}

The two scenarios share the same technical underpinnings, such as MHD BEV population, charger size, and kWh/mile driving efficiency. The principle difference between the two scenarios is the temporal distribution of charging availability, as demonstrated by the comparison of drive start times in Fig. \ref{fig:scenarios}.
By presenting both scenarios, it is possible to get a look at a larger picture of the range in potential cost savings of V1G and V2G and investigate the sensitivity with respect to the trip temporal patterns. These scenarios also raise additional questions regarding the total cost and savings associated with enabling these services. 

The HEVI-LOAD scenario (Scenario HL) relies on the results of the HEVI-LOAD tool, which \cite{HEVI-LOAD} is an agent-based driving and charging simulation tool for MHD zero-emission vehicles (ZEVs) developed by the Lawrence Berkeley National Laboratory in collaboration with the California Energy Commission (CEC). HEVI-LOAD takes multiple data sources as input and resolves the integrated driving, parking, and charging/refueling behaviors of the future MHD ZEVs. Individual trucks are referred to as agents whose behaviors are constructed and calibrated utilizing multiple data sources, such as adoption projection, travel demand, telematics data, power-train specifications, etc. Trip origin and destinations are provided at the traffic analysis zones (TAZ) level for better geospatial granularities. The overall trip statistics in terms of vehicle miles traveled (VMT), energy consumption rate (kWh/mile), and vehicle stock by segment have been validated with existing state policies. HEVI-LOAD creates a virtual environment that replicates real-world transportation scenarios with fine-grained representation of electrification scenarios. However, the high geospatial resolution that HEVI-LOAD charging profiles provide are obfuscated in this study to match the load zones as we consider only CAISO-level load. 

The additional scenario with varied temporal patterns (Scenario FD) is informed by the Fleet DNA dataset. This dataset is composed of thousands of historical drives across a variety of vehicle classes, vocations, and days. Each entry has several hundred associated fields, but for our purposes, the key information extracted is drive start time, drive end time, and VMT. Then, for each drive, the efficiency mapping in Table \ref{tab:evassumptions} is used to convert VMT to kWh consumption. We assume that each vehicle charges to 100\% before departing. The SoC at depot arrival can be calculated as the difference between the capacity and consumption. This dataset is combined with the California Energy Commission's 2023 AATE3 truck adoption projections \cite{AATE3}. Similarly to the approach in \cite{NREL_evpro}, we bootstrap from the Fleet DNA dataset according to the population projections by class and vocation. There are several key assumptions. Of course, bootstrapping assumes that the distribution of drive timing and distance present in Fleet DNA is the same as future MHD BEV drives in California. We assume that every vehicle drives and charges every day. It is also assumed that all charging occurs at the depot and there is no en-route charging. 
\begin{table}[h]
\caption{MHD BEV Technical Assumptions}
%\begin{center}
\centering
\begin{tabular}{ |c|c|c|c| } 
\hline
 & Charger Size & Capacity & Efficiency \\
 & (kW) & (kWh) & (kWh/mile) \\
 \hline
 Class 2-3 & 150 & 100 & 0.6  \\
\hline
 Class 4-6 & 150 & 300 & 1.05   \\
\hline
 Class 7 & 150 & 400 & 1.1  \\
\hline
 Class 8 & 150 & 600 & 1.8  \\
\hline
\end{tabular} \label{tab:evassumptions}
%\end{center}
\end{table}

As previously mentioned, modeling each vehicle individually would make computations intractable. For both scenarios, it is necessary to cluster the individual vehicles, and the same approach is used. We assume that if the vehicle is not driving, it is plugged in at the depot, and vice-versa. As dispatch is modeled hourly, vehicle charge start times are rounded to the next hour and vehicle charge end times are rounded to the previous hour. This rounding is to prevent an overestimation in the time flexibility of vehicles.
First, clusters are generated by enumerating all possible combinations of start and end hour. Each vehicle is assigned to a cluster. If the cluster size accounts for less than 0.1\% of all vehicles, this cluster is not modeled with V1G or V2G and left with a fixed charging profile, as this cluster would increase the associated complexity of the problem while only mildly impacting the solution due to the small number of associated controlled vehicles. This results in 87 clusters for Scenario HL, comprising in total 92\% of all vehicles and 168 clusters for Scenario FD, comprising in total 94\% of all vehicles. 
%For Scenario HL, fixed charging profiles are provided by HEVI-LOAD. For Scenario FD, fixed charging profiles are generated according to vehicles charging with the lowest rate of charge.

As a result of the assumptions made in Scenario FD, and the methodology of Scenario HL, the two scenarios have some key differences in addition to the trips' temporal patterns. In Scenario HL, approximately 1 out of 3 vehicles charge each day, as many vehicles make short trips and do not need to charge. Scenario FD does not account for this, and charges each vehicle daily. %This is an advantage of the agent-based simulation method, as Scenario FD does not account for such details.
However, because the underlying assumptions on VMT per day and truck efficiency are similar, the total daily MHD BEV load is extremely similar, within 1\%. 
This means in Scenario FD, the vehicles have considerably higher starting SoC, as well as a much larger number of vehicles connected resulting in considerably higher total power and energy capacity. The results will reflect this, and the ensuing discussion will consider both the pros and cons of this detail in terms of cost. 

We consider 3 charging regimes for both scenarios: a baseline case in which all charging is fixed, V1G, and V2G. 
For Scenario HL, fixed charging profiles are provided by HEVI-LOAD. %For Scenario FD, fixed charging profiles are generated according to vehicles charging with the lowest rate of charge.
For Scenario FD, the fixed charging profile is generated using the assumption that 50\% of vehicles charge immediately at full power and 50\% charge with the lowest power to fully charge by departure. For simplicity, all chargers are assumed to have 150kW rating.

\subsection{Results}
\begin{figure}[htb]
\centering
\includegraphics[width=0.9\linewidth]{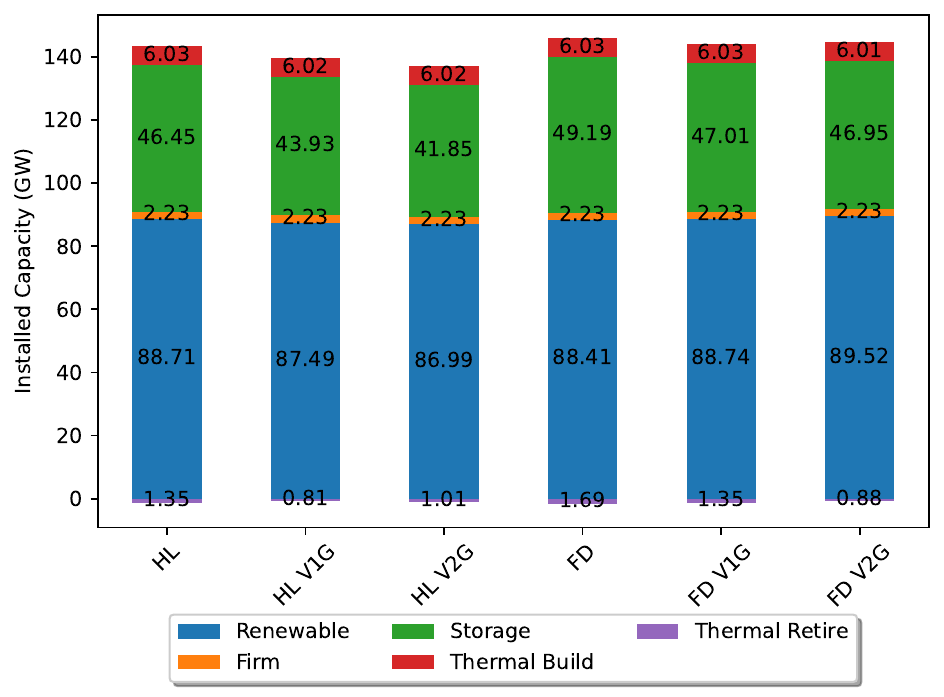}
\caption{Comparison of installed resources in 2045.}
\label{fig:fleet}
\end{figure}

The key consideration related to V1G and V2G with respect to decarbonization planning is quantifying how enabling these services lower the cost of power system decarbonization through lower investment, and potentially lower operation costs. Figure \ref{fig:fleet} shows the cumulative added capacity in year 2045. In general, V1G and V2G are associated with lower build of renewable and storage resources. 

By enabling V1G and V2G services, it is possible to avoid some of the installation of renewable and storage capacity that are needed in the base scenario to meet emissions targets. Accordingly, there are slightly less retirements of thermal units, which are kept online to meet the planning reserve margin.

\begin{figure}[htb]
\centering
\includegraphics[width=0.9\linewidth]{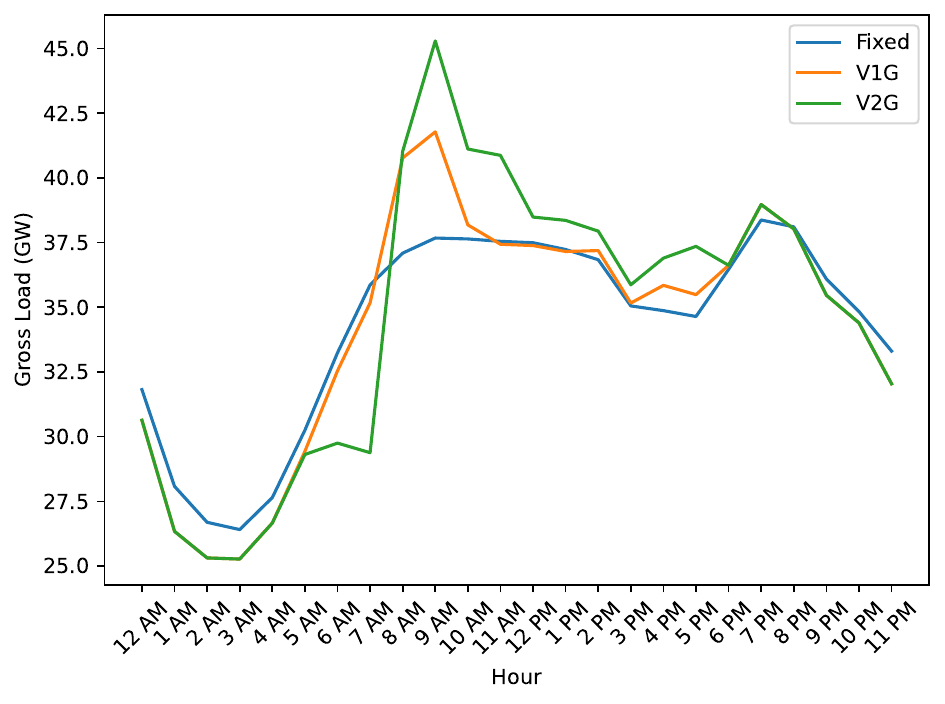}
\caption{Scenario FD gross load considering fixed charging, V1G, and V2G.}
\label{fig:grossload}
\end{figure}

\begin{figure}[htb]
\centering
\includegraphics[width=0.9\linewidth]{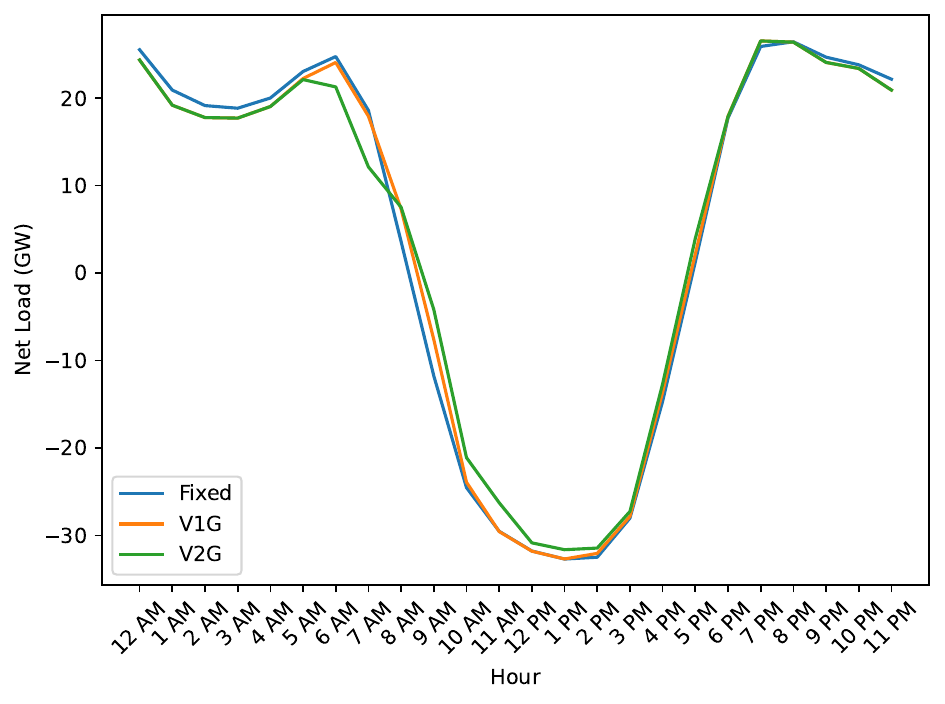}
\caption{Scenario FD net load for an exemplary day in 2035.}
\label{fig:netload}
\end{figure}

The mechanism by which these services lower investment costs is straightforward. Figure \ref{fig:grossload} shows the gross load for an exemplary day in 2035 under fixed charging, V1G, and V2G. Load is shifted from hours with lower renewable generation to hours with higher renewable generation. In the case of V2G, MHD BEV are able to provide power injections at critical hours to further reduce the need for energy storage. Most MHD BEV spend the bulk of the day driving, and thus are unable to charge when there would be most excess generation. As such, charging is mostly correlated to periods with lower variable renewable generation, and the cost savings comes mostly as avoided storage investment. This behavior is demonstrated by the visualization of net load for each regime in Fig. \ref{fig:netload}. V2G flattens the net load peak in the early morning, and recovers energy through the afternoon by charging batteries of stationary vehicles when renewable generation is plentiful. 

\begin{figure}[htb]
\centering
\includegraphics[width=0.9\linewidth]{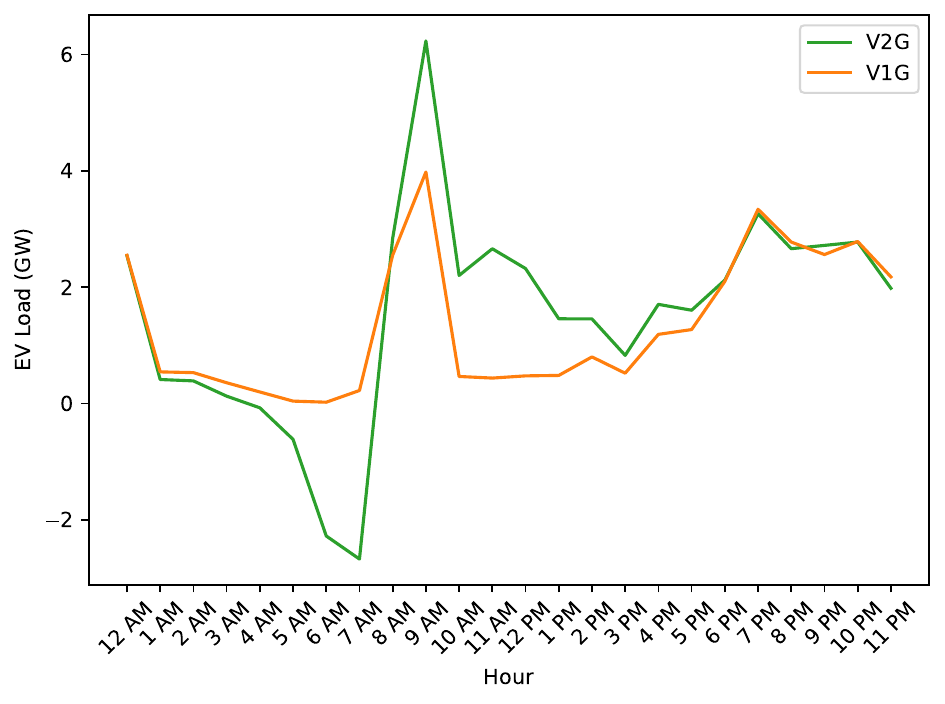}
\caption{Scenario FD MHD BEV hourly load averaged over year 2035.}
\label{fig:evload_ave}
\end{figure}

%Figure \ref{fig:evload} shows the MHD BEV load for an exemplary day in 2035 for V1G and V2G. 
Figure \ref{fig:evload_ave} shows the MHD BEV load for each hour, averaged over the year 2035.
The shape of V1G and V2G load is broadly similar, with the key difference that V2G is providing power to the grid for early morning hours, between 4am and 8am, then charging quickly between 8am and 10am, when the bulk of vehicles are leaving. 
%Figure \ref{fig:evload_ave} shows the MHD BEV load for each hour, averaged over the year 2035. 
There is a large spike in charging load in the morning, as other system loads are generally lower and solar generation ramps up. This spike is even larger for V2G, as the vehicles provide power in the very early morning. 
%\begin{figure}[h]
%\centering
%\includegraphics[width=1\linewidth]{ev_load.pdf}
%\caption{Scenario FD MHD BEV hourly load for an exemplary day.}
%\label{fig:evload}
%\end{figure}

The total costs as well as costs broken down by component are shown in Table \ref{tab:costs}.
As a resulted of the avoided investment in storage, there are substantially lower investment and maintenance costs. These services also help lower operational costs by lowering the use of thermal units. 
%Scenario HL primarily lowers costs by avoided investment and maintenance cost. 
Scenario HL V1G and V2G present 3.5\% and 4.6\% savings over baseline in California costs, respectively. 
%Scenario FD lowers costs through a more even balance of the three cost components.
Scenario FD V1G and V2G present 1.8\% and 3.0\% savings.
\begin{table}[h]
\caption{Costs, billions 2025\$}
\centering
%\begin{center}
\begin{tabular}{ |c|c|c|c|c|c|c| } 
\hline
   %& \multirow{2}{*}{Base} & \multicolumn{2}{c|}{Scenario HL} & \multicolumn{2}{c|}{Scenario FD} \\
   &  \multicolumn{3}{c|}{Scenario HL} &  \multicolumn{3}{c|}{Scenario FD} \\   
 \hline
 & Fixed & V1G & V2G & Fixed & V1G & V2G \\
 \hline
 Total Cost & 397.5 & 384.4 & 381.0 & 401.8 & 397.1 & 391.8 \\
 \hline
 CA Cost & 247.8 & 239.2 & 236.3 & 251.7 & 247.1 & 244.2 \\
 \hline
 Maint. Cost & 67.0 & 65.8 & 65.9 & 67.2 & 67.0 & 67.137 \\
 \hline
 Inv. Cost & 152.3 & 145.5 & 142.5 & 156.0 & 151.9 & 149.1 \\
 \hline
 CA Op. Cost & 28.5 & 27.9 & 27.9 & 28.4 & 28.2 & 28.0 \\
 \hline
\end{tabular} \label{tab:costs}
%\end{center}
\end{table}

The cost savings of V1G and V2G over fixed charging are shown per vehicle, per year in Table \ref{tab:levelizedcosts}. These costs are not discounted for the time value of money. In the worst case, V1G saves a few hundred dollars per vehicle per year. In the best case, V2G saves several thousand dollars for each vehicle each year. 
\begin{table}[h]
\caption{Levelized Cost Savings over Fixed Charging (\$ per vehicle-year, Non-discounted)}
%\begin{center}
\centering
\begin{tabular}{ |c|c|c|c|c|c| } 
\hline
 & 2025 & 2030 & 2035 & 2040 & 2045 \\
 \hline
 HL V1G & 2765 &  1337 & 1243 & 1378 & 997 \\
 \hline
 HL V2G & 4317 & 1840 & 1510 & 1822 & 1350 \\
 \hline
 FD V1G & 871 & 592 & 431 & 457 & 724 \\
 \hline
 FD V2G & 1277 & 1204 & 933 & 1005 & 999 \\
 \hline
\end{tabular} \label{tab:levelizedcosts}
%\end{center}
\end{table}
\subsection{Policy Implications}
The decision to enable V1G and V2G services does not exist in a vacuum, and it is crucial to quantify potential costs related to these services. The main two considerations are the cost of battery degradation and the cost of charging infrastructure.

\subsubsection{Degradation}
Battery degradation is quantified using the BLAST model \cite{BLAST}. This model takes an input SoC time series and returns a total degradation \%. We run this model for each 5 year investment interval. The goal is to understand how V1G and V2G services impact battery degradation over default operation. Understanding how the batteries degrade over this interval helps evaluate the overall cost and value of these services. 

Each cluster of MHD BEV is evaluated for degradation independently. 
The SoC time series is created by stacking the MHD BEV SoC time series of each representative period by their respective weights to make a yearly time series, then stacking that time series to obtain a 5-year-long time series corresponding to the investment frequency. We calculate the degradation given 3 battery chemistries (lithium-iron-phosphate, nickel-cobalt-aluminium, and nickel-manganese-cobalt) and take the average as the final degradation \%. This percentage can then be converted to a total degraded kWh given the kWh capacity of each cluster. 

Degradation cost is estimated by making the assumption that, at the end of the interval, batteries can be refurbished by replacement of cells to restore battery health. Thus, cost is linear with degradation. Of course, the cost of battery degradation is more complex than this, but this method provides an effective way of comparing the relative degradation between BEV charging regimes and scenarios. %In particular, we are interested in the comparison between V1G and V2G. V1G and the baseline charging profile should experience similar levels of degradation, because no cycling of batteries occurs and they are subject to the same limit on charge rate. 

We assume a battery degration cost of \$100/kWh. In 2022, the cost of battery packs reached ~\$150/kWh \cite{doebatterycost}. The cost of battery packs are expected to drop further, with projections covering a significant range. \cite{nrelbatterycost} predicts grid stationary battery costs will see a reduction of 16\% to 47\% by 2030. \cite{bloomberg} estimates a lithium-ion battery pack cost of 72\$/kWh (in 2022 \$) by 2030. Thus, \$100/kWh should be fairly conservative. 

We examine the degradation for the Scenario FD. We present degradation for a base scenario, V1G, and V2G. 
%In the baseline scenario, the vehicles charge with the minimum power profile. 
Scenario FD is a good candidate for quantifying degradation because each vehicle is controlled. There is not a rigorous way of measuring degradation in Scenario HL, because during the optimization, roughly 1 in 3 vehicles charge each night. From the vehicle perspective, some MHD BEV are charging every night and some are charging less frequently. From the perspective of the grid, it does not matter which vehicles are plugging in. As a consequence, this does not permit rigorous tracking of each vehicle's SoC. 

The cost of degradation as well as the average relative battery capacity at the end of each 5-year interval is shown in Table \ref{tab:deg}.
The impact of degradation is relatively mild. The vast majority of the degradation seems to be due to aging.
%Due to seasonal temperature fluctuations, the rate of degradation is variable through the year, as shown by the oscillations in Fig. \ref{fig:deg}. 
Batteries experience on average an extra 0.2\% of degradation for V1G vs the baseline case, and an additional 0.3\% again for V2G. The critical consideration is the increase in degradation costs over baseline. Operating vehicles will necessarily incur degradation, but it is critical to understand what costs are incurred by V1G and V2G services. The cost associated with degradation is increased by 0.1 billion USD for V1G and 0.2 billion USD for V2G, as compared to the baseline. Although these costs are considerable, they are an order of magnitude less than the potential savings. As such, increased degradation is a relevant consideration, but it is not a critical risk to the business case for V1G and V2G services.
\begin{table}[h]
\caption{Battery degradation}
\centering
%\begin{center}
\begin{tabular}{ |c|c|c|c| } 
\hline
 & Baseline & V1G & V2G \\
 %\hline
 \hline
 %Degradation Cost (Billions) & 4.7 & 4.9 & 5.2\\
 Degradation Cost (Billions) & 7.5 & 7.6 & 7.7\\
 \hline
 %Average 5-Year Degradation \% & 82.5 & 81.6 & 80.4 \\
 Residual Discharge Capacity \% & 81.9 & 81.7 & 81.4 \\
 \hline
\end{tabular} \label{tab:deg}
%\end{center}
\end{table}

\subsubsection{Cost of Chargers}
In terms of BEV supply equipment costs, the most relevant factors are the cost of ensuring vehicles have sufficient access to chargers, and the cost of enabling bidirectional charging over unidirectional charging. 

At time of writing, there are very limited number of V2G ready chargers on the market. Bidirectional chargers are substantially more expensive that unidirectional chargers, but it is difficult to estimate how much of that cost difference is driven by the lack of commercialization. While numerous studies examine the cost of BEV supply equipment, there are no concrete comparisons of the cost of bidirectional and unidirectional MHD BEV supply equipment. To estimate the potential cost of bidirectional chargers vs unidirectional chargers, we consider two elements which are necessary for enabling bidirectional charging. The first is an islanding switch, which can be opened to prevent energy flowing into lines, for example, when lines must be serviced. The cost of this switch is likely negligible if it is installed at the time that the charging depot is constructed. The other cost is an inverter required to convert the DC current of the MHD BEV battery to AC used by the grid. We estimate this cost using the cost of solar inverters, approximately \$50 per kW \cite{NREL_solar_inverter}. The total cost of this equipment adds \$1.1B to the V2G cost of Scenario HL in Table \ref{tab:costs}. These costs reduce substantially the potential savings of V2G. We should emphasize that the upcharge associated with V2G is purely speculative. Depending on the cost of bidirectional equipment, V2G could pose a better or worse business case. 

%As demonstrated above, daily charging in Scenario FD does present substantial savings from the power system perspective. However, Scenario HL demonstrates that it is possible to enable charging and driving for these vehicles without having one vehicle per charging. The goal of the analysis here is to quantify the cost of enabling all vehicles to charge each day. 

%As demonstrated above, Scenario FD presents slightly higher cost savings for V2G over V1G than Scenario HL. The likely reason for this is that, by charging daily, vehicles generally enter the depot with a higher state of charge. This frees up more energy for 

The two scenarios are generated under different basic charging behavior assumptions, and these assumptions impact the cost related to charging in a major way.
Scenario HL is an agent-based approach, in which vehicles only charge when necessary. %Many vehicles make short trips each day, and as a result, do not need to charge on a daily basis. 
As such, approximately 1 in 3 vehicles charge on a given day, and the number of chargers can be provided accordingly.
A key assumption of Scenario FD is that each vehicle charges each day. 
We consider two cases which bookend the spectrum on which this could be enabled. The first is providing every vehicle in Scenario FD with a dedicated charger. The cost of a 150kW DC fast charger is estimated at \$142,200 for hardware and installation \cite{crc}. For each investment interval, we calculate the cost of installing a dedicated charger for each vehicle in Scenario FD and installing only the necessary chargers in Scenario HL. In Scenario HL, we assume that a dedicated charger is installed for each vehicle charging in a given day. In total, the cost of chargers in Scenario HL would be \$20.6B and \$61.6B for Scenario FD.
%The additional cost of installing one charger per vehicle is \$31B. 
The second is providing only the necessary number of chargers. An emerging concept is to connect multiple vehicles to a single charger. If a charger is rated at 150kW, it may be able to connect to multiple vehicles simultaneously and provide either lower power to all, or full power to individual vehicles at different times. This service could be enabled without performing substantial hardware upgrades, only by providing some additional switchgear and plugs. 
If we take inspiration from this, we can suggest that in Scenario FD, the number of chargers needed is proportional to the peak hourly charging demand. This brings the number of necessary chargers down substantially, to approximately 1 charger per 4 vehicles in most years. Accordingly, the cost of installing chargers drops to \$14.5B.
Although installing chargers is essential with or without V1G and V2G, the range in potential charger costs is extremely large, and is bigger than the potential savings associated with these services.
%The range of cost associated with installing chargers in Scenario FD is \$14.5B in to \$61.6B, and Scenario HL is associated with charger costs of \$20.6B. 
Because of this, minimizing the number of necessary chargers is a very relevant consideration alongside lowering chargers costs with V1G and V2G.

%The savings for V1G Scenario FD vs Scenario HL are very similar, and this cost of additional chargers would dwarf potential savings in V1G. For V2G, the savings are very substantial for Scenario FD over Scenario HL; however, these savings would still be essentially wiped out by the cost of installing additional charging infrastructure. 

%\textcolor{blue}{If we want to include the upcharge of bidirectional chargers vs unidirectional chargers, there are two key components: inverter and islanding switch. Islanding switch is likely negligible because it would be built at the depot level, and should have negligible install cost if installed with the charger. Inverter can be estimated at the price of current solar inverter cost, NREL says about \$50 per kW. so \$7500 per charger in this scenario. Would the price drop if a rectifier (sunk cost)/inverter were integrated?}

\section{Conclusion} \label{sec6}
In this paper, we examined the potential costs and savings of enabling V1G and V2G services for MHD BEVs in California. 
Using a large scale MILP model, we calculate the savings of these services from the perspective of a central power system planner.
Two scenarios are used to understand the driving and charging behavior of vehicles.
We also estimate costs linked to these services. We show that battery degradation is not insignificant, but is associated with costs an order of magnitude lower than potential savings. We estimate that the cost of enabling bidirectional charging could be a very relevant element, and could weaken the business case of V2G over V1G. Carefully identifying the number of necessary chargers is of utmost importance, as costs associated with chargers could be very large.

\bibliographystyle{ieeetr}
\bibliography{references}

\end{document}